\documentclass[preprint,notoc]{JHEP3}
\usepackage{amssymb,amsmath,bbm}
\usepackage{epsfig}
 
\DeclareOldFontCommand{\rm}{\normalfont\rmfamily}{\mathrm}
\DeclareOldFontCommand{\bf}{\normalfont\bfseries}{\mathbf}
\def\tz{\widetilde Z}
\def\tw{\widetilde W}
\def\ttau{\tilde \tau}
\def\tnu{\tilde\nu}
\def\ttop{\tilde t}
\def\tL{\tilde L}
\def\tH{\tilde H}

\def\mns{{\bf m}^2_{\tnu_R}}

\def\bfu{{\bf f}_u}

\def\bfn{{\bf f}_{\nu}}
\def\banu{{\bf a}_{\nu}}
\def\bbnu{{\bf b}_{\nu}}
\def\ftau{f_{\tau}}
\def\fnu{f_{\nu}}
\def\atau{A_{\tau}}
\def\anu{A_{\nu}}
\newcommand{\beq}{\begin{equation}}
\newcommand{\eeq}{\end{equation}}
\def\bea{\begin{eqnarray}}  
\def\eea{\end{eqnarray}}  

\title{Neutrino sector impacts SUSY dark matter}
\author{Vernon Barger$^{a,c}$, Danny Marfatia$^{b,c}$ and Azar Mustafayev$^{b,c}$\\
$^a$Department of Physics, University of Wisconsin, Madison, WI 53706, USA\\
$^b$Department of Physics \& Astronomy,
University of Kansas, Lawrence, KS 66045, USA\\
$^c$Kavli Institute for Theoretical Physics, University of California, Santa Barbara, CA~93106, USA\\
E-mail: \email{barger@physics.wisc.edu}, \email{marfatia@ku.edu}, \email{amustaf@ku.edu}}

\preprint{NSF-KITP-08-30}

\abstract{Motivated by the fact that neutrinos are massive, we study the effect of neutrino Yukawa couplings 
on neutralino dark matter observables within the framework of a supersymmetric seesaw. We find that neutrino couplings
significantly affect the neutralino relic density in regions of parameter space where
soft SUSY-breaking slepton masses and/or trilinear couplings are large. 
Depending on the size of the couplings, the neutralino relic density 
spans over an order of magnitude in the $A$-funnel, 
focus point and stop-coannihilation regions of mSUGRA. We also show that dark matter 
detection rates can be modified by up to several orders of magnitude. \\

\begin{center}
\end{center}

}

\begin{document}

\section{Introduction}

A very attractive aspect of $R-$parity conserving supersymmetry
(SUSY)~\cite{Baer:book,Drees:book} is that it predicts the existence of
a massive, electrically and color neutral, stable weakly interacting particle 
that is a natural Cold Dark Matter (CDM) candidate.
In most cases this particle is the lightest neutralino, $\tz_1$, whose relic abundance can be reliably calculated as a
function of model parameters~\cite{griest}.
Cosmological measurements determine the mass density of CDM with high
accuracy. A combination of WMAP data with distance measurements from the 
baryon acoustic oscillations in galaxy power spectra gives~\cite{wmap} 
\beq
\Omega h^2 = 0.1120^{+0.0074}_{-0.0076} \ \ \ \ (2\sigma)\;,
\label{eq:relic}
\eeq
where $\Omega\equiv \rho/\rho_c$ with $\rho_c$, the critical mass density of the
Universe, and $h$ is the scaled Hubble parameter. Such a precise measurement puts
severe constraints on new physics scenarios. 
For example, in the well-studied minimal supergravity model, mSUGRA (or CMSSM)~\cite{msugra}, 
the only surviving regions of parameter space are those in which neutralino annihilation is enhanced:
the bulk region~\cite{bulk,Baer:1995nc}, the stau~\cite{dm:stau,isared} or stop~\cite{dm:stop}
coannihilation regions, the hyperbolic branch/focus point (HB/FP) region~\cite{dm:fp}, and
the $A$ or $h$ resonance annihilation (Higgs funnel) regions~\cite{Baer:1995nc,Afunnel,hfunnel}.
The narrowness of these regions in mSUGRA motivated studies with one and two
parameter extensions, in which non-universality of soft SUSY-breaking parameters reduce $\Omega_{\tz_1}h^2$
in accord with WMAP data over a large portion of parameter space~\cite{wtn}. 

Other new physics that must be incorporated into the MSSM is that neutrinos are massive, as
indicated by the observation of neutrino ocillations~\cite{bmw}.
An elegant explanation for neutrino mass is offered by the
seesaw mechanism~\cite{seesaw1}. Here three right-handed neutrinos (RHNs) with large Majorana masses are added to the SM. 
The decoupling of these RHNs naturally generates light masses for the left-handed neutrinos. 
We demand that the MSSM accommodate neutrino masses via the seesaw mechanism and explore whether favorable
relic neutralino densities ensue by using the mSUGRA framework as a specific example. 
(Incidentally, this is potentially a more compelling scheme to
 expand the allowed mSUGRA parameter space than the introduction of non-universal soft SUSY-breaking parameters.)
We demonstrate the invalidity of the lore that 
neutrino Yukawa couplings below the Grand Unification Scale (GUT)
do not significantly affect Dark Matter (DM) observables.\footnote{Recently it 
was shown in the context of particular SUSY-GUT models that RHNs significantly impact low-energy
phenomenology~\cite{Calibbi}. However, the result is dominated by additional contributions to right-handed
slepton mass evolution {\it above} the GUT scale, which is highly model dependent. In this work we take a model-independent approach 
and study the effect on DM predictions due to large neutrino Yukawa couplings {\it below} the GUT scale.}

The rest of the paper is organized as follows. We briefly review the SUSY-seesaw mechanism in
section~\ref{sseesaw}. In section~\ref{seesaw2rd}, we study the effect of neutrino Yukawa couplings on DM
observables using benchmark points from mSUGRA augumented with RHN superfields (mSUGRA+RHN).
We show that the presence of large neutrino Yukawa couplings significanty affect the evolution of
sparticle masses with concomitant changes in DM observables. 
Finally, in section~\ref{disc} we discuss the impact on the mSUGRA+RHN parameter
space and on various non-universal models, and summarize our results.

\section{SUSY-seesaw}
\label{sseesaw}

The superpotential for the MSSM with right-handed neutrinos (MSSM+RHN), in the notations and conventions of Ref.~\cite{Baer:book}, 
can be written as
\beq
\hat{f}=\hat{f}_{MSSM}+
(\bfn)_{ij}\epsilon_{ab}\hat{L}^a_i\hat{H}_u^b\hat{N}^c_j +
\frac{1}{2}(M_N)_{ij}\hat{N}^c_i \hat{N}^c_j \, ,
\label{eq:superpot}
\eeq
where $\hat{f}_{MSSM}$ is the MSSM superpotential, $\hat{L}$ and $\hat{H}_u$ are, respectively, lepton doublet and
up-higgs superfields, 
$\hat{N}^c_i$ is the superfield whose fermionic component is the left-handed antineutrino and
scalar component is $\tnu^\dagger_R$, $i,j=1,2,3$ are generation indices, $a,b$ are $SU(2)_L$ doublet indices, 
$\bfn$ is the neutrino Yukawa coupling matrix and 
$M_N$ represents the (heavy) Majorana mass matrix for the right-handed neutrinos.
After electroweak symmetry breaking, Eq.~(\ref{eq:superpot}) leads to the the following mass matrix for the light neutrinos:
\beq
\mathcal{M}_{\nu} = \bfn M_N^{-1} \bfn^T v^2_u \equiv \kappa v^2_u \, ,
\eeq
where $v_u$ is the vacuum expectation value of the up-type Higgs field $h^0_u$.
This corresponds to the so-called type I seesaw~\cite{seesaw1}, where there are no other contributions 
({\it i.e.}, type II or type III) to the light neutrino masses.

Additional soft SUSY breaking (SSB) terms are included so that the Lagrangian becomes
\beq
\mathcal{L} = \mathcal{L}_{MSSM} - \tnu^{\dagger}_{Ri} (\mns)_{ij} \tnu_{Rj}
  +\left[ (\banu)_{ij}\epsilon_{ab} \tL_i^a \tH_u^b \tnu^{\dagger}_{Rj} +\frac{1}{2}(\bbnu)_{ij}\tnu_{Ri}\tnu_{Rj} + {\rm h.c.} \right] \, .
\label{eq:ssb}
\eeq
The parameters $\sqrt{(\mns)_{ij}},\ (\banu)_{ij}$ and $(\bbnu)_{ij}$ are assumed to be of order the weak scale, 
even though right-handed neutrinos and their superpartners have much larger masses $M_N$.

Without further assumptions the neutrino Yukawa couplings are arbitrary. 
We can estimate the size of $\bfn$ by turning to $SO(10)$ GUTs. Here, all SM fermions and 
the right-handed neutrino of each generation are unified in a single spinorial representation, $\bf{16}$,
of the $SO(10)$ gauge group. The product of two $\bf{16}$'s, that appear in the Yukawa term of a superpotential, 
can only couple to $\bf{10},\ \bf{120}$ or $\bf{126}$. 
If higgs superfields reside in $\bf{10}$, as in the simplest models, then $\bfn = \bfu$ at $M_{GUT}$; 
if the higgses occupy $\bf{126}$, then $\bfn =3\bfu$. Contributions from $\bf{120}$ can only be subdominant, since they
would lead to at least a pair of degenerate heavy up-quarks~\cite{so10}.

In the renormalization group equations (RGEs), the Yukawa matrix appears as $\bfn^\dagger \bfn$ and 
is dominated by the (3,3) element; if $\bfn = \bfu$ then $(\bfn^\dagger \bfn)_{33} \simeq 0.25$ at $M_{GUT}$,
while the other elements are smaller by two to five orders of magnitude.
Also, the off-diagonal entries in the SSB
matrices at the weak scale have to be very small to meet stringent flavor-violation constraints and consequently they
do not affect the mass spectrum significantly~\cite{fcnc}. 
Therefore, for the sake of clarity, we present our discussion in the third-generation dominant scheme, {\it i.e.}, we
assume ${\bf f}_{u,d,e,\nu} \sim \left({\bf f}_{u,d,e,\nu}\right)_{33} \equiv f_{t,b,\tau ,\nu}$,
$M_N \sim \left(M_N \right)_{33} \equiv M_{N_3}$, 
${\bf a}_{u,d,e,\nu} \sim \left({\bf a}_{u,d,e,\nu}\right)_{33} \equiv -A_{t,b,\tau ,\nu}\, f_{t,b,\tau ,\nu}$
and the SSB mass squared matrices are diagonal at all scales.
However, the numerical analysis is performed in the full matrix form.\footnote{We assume a 
normal hierarchy for $\bfn$ and $M_N$ for consistency with the other SM fermions.
This does not restrict the light neutrino mass eigenstates to have a normal hierarchy.}

The right-handed neutrino superfields above the seesaw scale modify the evolution of gauge $g_1,\ g_2$ coupling at 
the 2-loop level, charged lepton and up-quark Yukawa couplings at the 1-loop level and introduce new RGEs
for neutrino Yukawa couplings and Majorana masses~\cite{rges}. 
The RGEs for the parameter $\bbnu$ are irrelevant for our analysis.
Among the RGEs for the SSB parameters, the following get modified at the 
1-loop level~\cite{casas,baer:rges} and are relevant for our discussion:
\bea
\frac{d m^2_{L_3}}{dt} &=& \frac{2}{16\pi^2}\left\lbrack -\frac{3}{5}g_1^2 M_1^2 -3g_2^2 M_2^2 -\frac{3}{10}g_1 S
                 +\ftau^2 X_{\tau} +\fnu^2 X_{\nu} \right\rbrack \label{eq:ml3}\\
\frac{d m^2_{\tnu_R}}{dt} &=& \frac{4}{16\pi^2} \fnu^2 X_{\nu} \\
\frac{d \atau}{dt} &=& \frac{2}{16\pi^2}\left\lbrack \sum c_i'' g_i^2 M_i 
                 +3f_b^2 A_b +4\ftau^2 \atau +\fnu^2  \anu \right\rbrack \label{eq:atau}\\
\frac{d A_t}{dt} &=& \frac{2}{16\pi^2}\left\lbrack \sum c_i g_i^2 M_i 
                 +3f_t^2 A_t +f_b^2 A_b +\fnu^2 \anu \right\rbrack \label{eq:at}\\
\frac{d m^2_{H_u}}{dt} &=& \frac{2}{16\pi^2}\left\lbrack -\frac{3}{5}g_1^2 M_1^2 -3g_2^2 M_2^2 +\frac{3}{10}g_1 S
                 +3f_t^2 X_t +\fnu^2 X_{\nu} \right\rbrack  \label{eq:mhu}
\eea
where  $t=\log Q,\ X_{\nu}=m^2_{L_3}+m^2_{\tnu_R}+m^2_{H_u}+\anu^2$, and $X_t,\ X_b,\ X_{\tau},\ S$ and $c_i ,\ c_i''$ are given
in Ref.~\cite{baer:rges}.
The rest of the RGEs remain unchanged at the 1-loop level from those for the MSSM and may be found in Ref.~\cite{martin}. 
Below the seesaw scale the right-handed neutrinos are integrated out and the MSSM is the effective theory with
an appropriate change of RGEs; the only remaining trace is in the evolution of the dimension-5 neutrino operator $\kappa$~\cite{rges}.

Usually, the $f_t^2 X_t$ term dominates over gauge-gaugino terms and causes radiative electroweak symmetry breaking (REWSB)
by driving $m^2_{H_u}$ to negative values. In the case of the MSSM+RHN, $m^2_{H_u}$ is driven to more negative values
by the $\fnu^2 X_{\nu}$ term.
Similarly, the third generation slepton doublet mass $m^2_{L_3}$ is driven to smaller values in the MSSM+RHN by the $\fnu^2 X_{\nu}$ term. 
Unless the trilinear A-terms have very large GUT-scale values, they are pushed to negative values by the gauge-gaugino terms.

These features are illustrated in Fig.~\ref{fig:evol}, which shows the running of third generation slepton
doublet, up- and down-higgs mass parameters  
and A-terms from $M_{GUT}$ to $M_{weak}$ for mSUGRA (solid curves) and two cases of mSUGRA+RHN 
with different values of GUT-scale neutrino Yukawa coupling $f_{\nu}$ (dashed and dotted 
curves). 
\FIGURE{
\epsfig{file=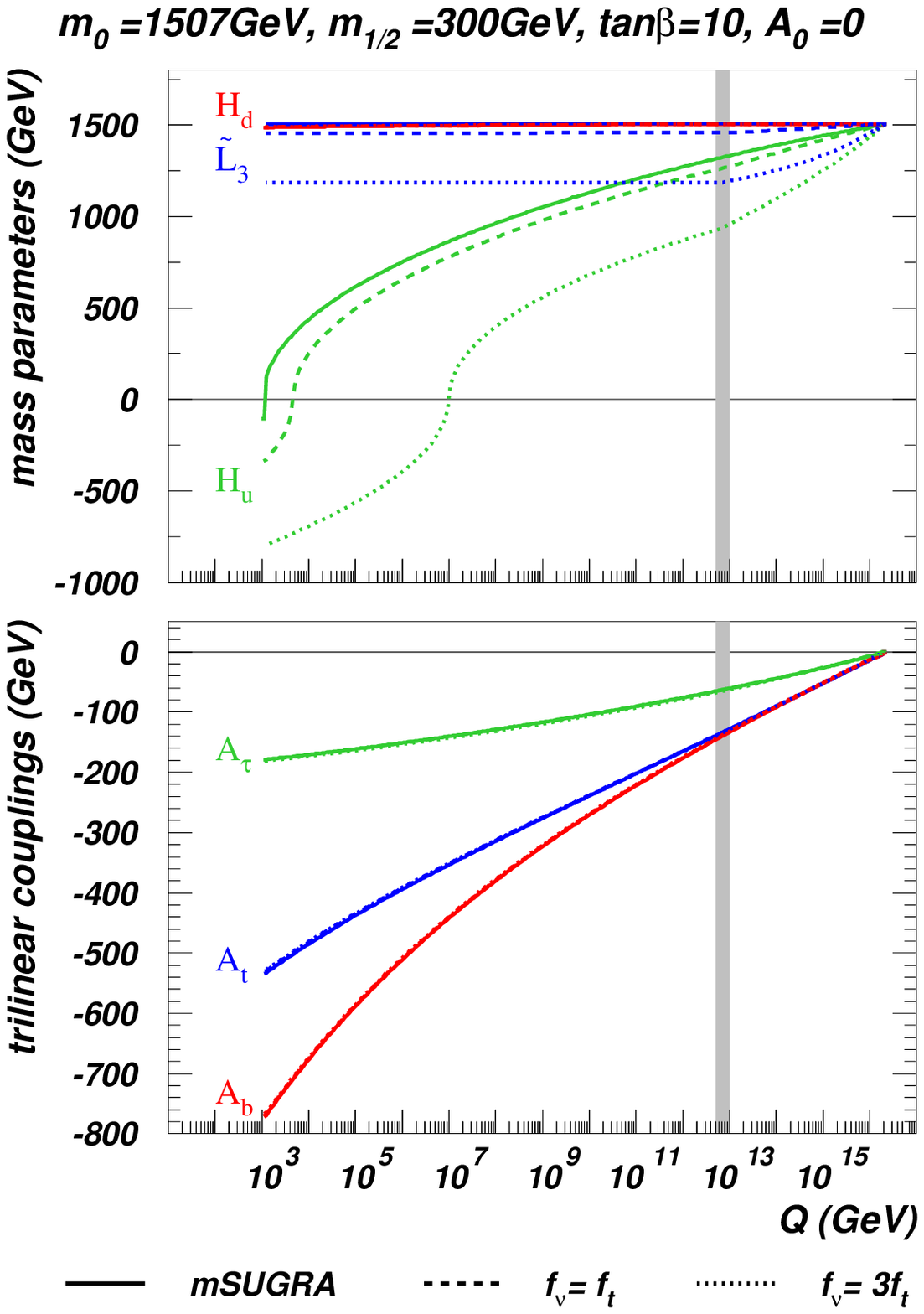,width=7.4cm,angle=0} 
\epsfig{file=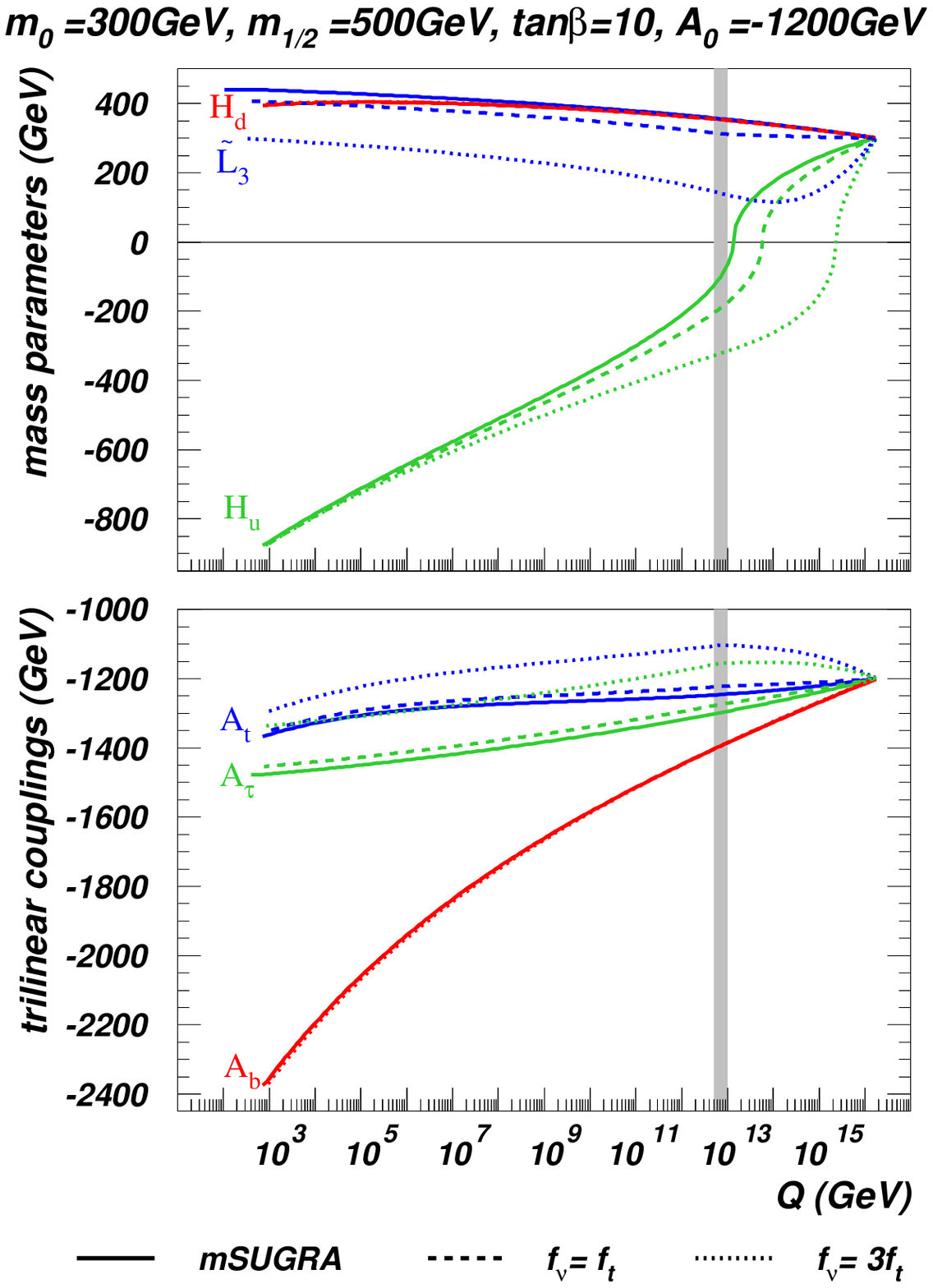,width=7.4cm,angle=0} 
\caption{\label{fig:evol} 
Evolution of the soft SUSY breaking
Higgs mass parameters ${\rm sgn}(m_{H_u}^2) \sqrt{|m_{H_u}^2}|$ 
and ${\rm sgn} (m_{H_d}^2) \sqrt{|m_{H_d}^2}|$, the third generation slepton doublet 
 mass $m_{\tL_3}$, and trilinear couplings as a function of the
renormalization scale $Q$ in the mSUGRA model (solid), and two mSUGRA+RHN models --
with $\fnu(M_{GUT})=f_t(M_{GUT})$ (dashed lines) and 
with $\fnu(M_{GUT})=3f_t(M_{GUT})$ (dotted lines).
The model parameters are $m_0=1507$~GeV, $m_{1/2}=300$~GeV,
$A_0=0$ for the left frames and  
$m_0=300$~GeV, $m_{1/2}=500$~GeV, $A_0=-1200$~GeV for the right frames. 
We take $\tan\beta =10$, $\mu >0$ and $m_t=171$~GeV and also fix $M_{N_3} = 10^{13}$~GeV for mSUGRA+RHN models.
Grey bands indicate the scale at which the third generation right-handed neutrino decouples.}
}
In the left frames we consider the case in which large $X_{\nu}$ values result from large scalar masses.
The trilinear A-terms evolve identically for the three models because the smallness of the A-terms at the GUT scale
nullifies the upward push of $\fnu$.
As expected, $m^2_{H_u}$ and $m^2_{L_3}$ run to smaller values for mSUGRA+RHN. 
We see kinks in their evolution at the scale $Q \sim M_{N_3}$, where the right-handed neutrino
decouples and the $\fnu^2 X_{\nu}$ and $\fnu^2 \anu$ terms do not contribute to the RGEs. 
Since right-handed neutrinos do not couple to the down-higgs directly, the RGE for $m^2_{H_d}$ is unaffected.
In the right frames we consider the case in which large $X_{\nu}$ values result from large trilinear A-terms.
 Now, the effect of $\fnu$ in Eqs.~(\ref{eq:atau}-\ref{eq:at}) is non-negligible as is evident from the
smaller absolute values for $\atau$ and $A_t$ at the weak scale. As in the previous case, the large $X_{\nu}$ 
pushes the slepton doublet mass to
smaller values. However, the picture is different for the up-higgs mass. Initially, the large
$\anu$ contribution to $X_{\nu}$ pushes $m^2_{H_u}$ to more negative values. This reduces the magnitudes of $X_{\nu}$ and $X_t$, 
thus increasing the role of the gauge-gaugino terms. Consequently, weak-scale values of $m^2_{H_u}$ are almost unchanged from
those in mSUGRA.

For moderate to large $\tan\beta$ values, as favored by the Higgs boson mass constraint from LEP2, 
the tree-level\footnote{We use tree-level relations only for illustration. In practice, $\mu^2$ and sparticle spectrum are
determined using the full 1-loop expressions.} 
minimization condition for EWSB in the MSSM 
can be approximated as $\mu^2 \simeq -m^2_{H_u} -0.5M_Z^2$, and the  
 the tree-level CP-odd Higgs boson mass is 
$m_A^2 \simeq m^2_{H_d} - m^2_{H_u}-M_Z^2$, which implies that when $|m^2_{H_u}|$ is large, 
both $|\mu|$ and $m_A$ are large and correspondingly the $H$ and
$H^{\pm}$ Higgs bosons are heavier.  
Also note that 
larger weak-scale A-terms lead to more mixed states and a lighter $\tilde{f}_1$ sfermion state.

\section{Seesaw and Relic Density}
\label{seesaw2rd}

In the MSSM, the neutralino is a superposition of the bino, wino, up- and down-higgsino states.
Over the most of the mSUGRA parameter space the lightest neutralino $\tz_1$ is mainly bino with a 
small annihilation cross section so that its relic density is 
significantly above the WMAP range. 
Nevertheless, there exist regions of parameter space
where various mechanisms enhance the annihilation and WMAP bounds can be satisfied.
To examine effects of a SUSY-seesaw on DM observables in these regions, 
we use the mSUGRA model as an example.
The model is completely specified by the well-known parameter set
$m_0,\ m_{1/2},\ A_0,\ \tan \beta ,\ {\rm sgn}(\mu)$.
The SSB terms are taken to be universal at the GUT scale, $M_{GUT}\simeq 2 \times 10^{16}$~GeV, 
with $m_0$ the common scalar mass, $m_{1/2}$ the common gaugino mass, and $A_0$
the common trilinear coupling.
The universal SSB boundary conditions allow us to isolate the neutrino Yukawa coupling effects.
Since electroweak symmetry is broken radiatively, the
magnitude (but not the sign) of the superpotential Higgs mass term $\mu$ is determined, 
and we can trade the GUT-scale bilinear
soft term $B$ for the weak-scale ratio of Higgs vacuum expectation values, $\tan \beta$. Once
weak-scale values of the SSB parameters are computed via renormalization group evolution, 
they serve as inputs for computation of sparticle masses and mixings. Then,
the neutralino relic density $\Omega_{\tz_1}h^2$ and DM detection rates can be calculated.

In each DM-allowed region of mSUGRA we select one benchmark point\footnote{These points 
are closely related to SPS benchmark points~\cite{sps}. The small difference in parameter
values is attributable to the different code and top mass we are using.} from Table~\ref{tab:points}
 and vary $M_{N_3}$. 
The values of the first and second generation right-handed neutrino masses are not important, 
since the corresponding Yukawa couplings are assumed to be
very small. For definiteness, we fix their masses to $10^{11}$~GeV and $10^{12}$~GeV, respectively.
For the GUT-scale neutrino Yukawa coupling we consider the two cases introduced earlier: 
$\fnu (M_{GUT})=f_t(M_{GUT})$ and $\fnu (M_{GUT})=3f_t(M_{GUT})$.
We take $\mu >0$ as suggested by experimental measurements of the muon anomalous magnetic moment~\cite{g-2}, 
and top mass $m_t = 171$~GeV to conform with Tevatron data~\cite{mtop}.
\TABLE{
\begin{tabular}{c||cccc|c}
\hline
Point & $m_0$ & $m_{1/2}$ & $A_0$ & $\tan\beta$ & Region \\
\hline
A &  80 & 170 & -250 & 10 & bulk \\
B & 100 & 500 & 0 & 10 & $\ttau$-coan.\\
C & 150 & 300 & -1091 & 5 &  $\ttop$-coan.\\
D & 500 & 450 & 0 & 51 & $A$-funnel \\
E & 1507 & 300 & 0 & 10 & HB/FP \\
\hline
\end{tabular}
\caption{\label{tab:points} 
Input parameters for benchmark points and corresponding DM-allowed regions of mSUGRA. For
all points $\mu >0$ and $m_t = 171$~GeV.}
}

\FIGURE{
\epsfig{file=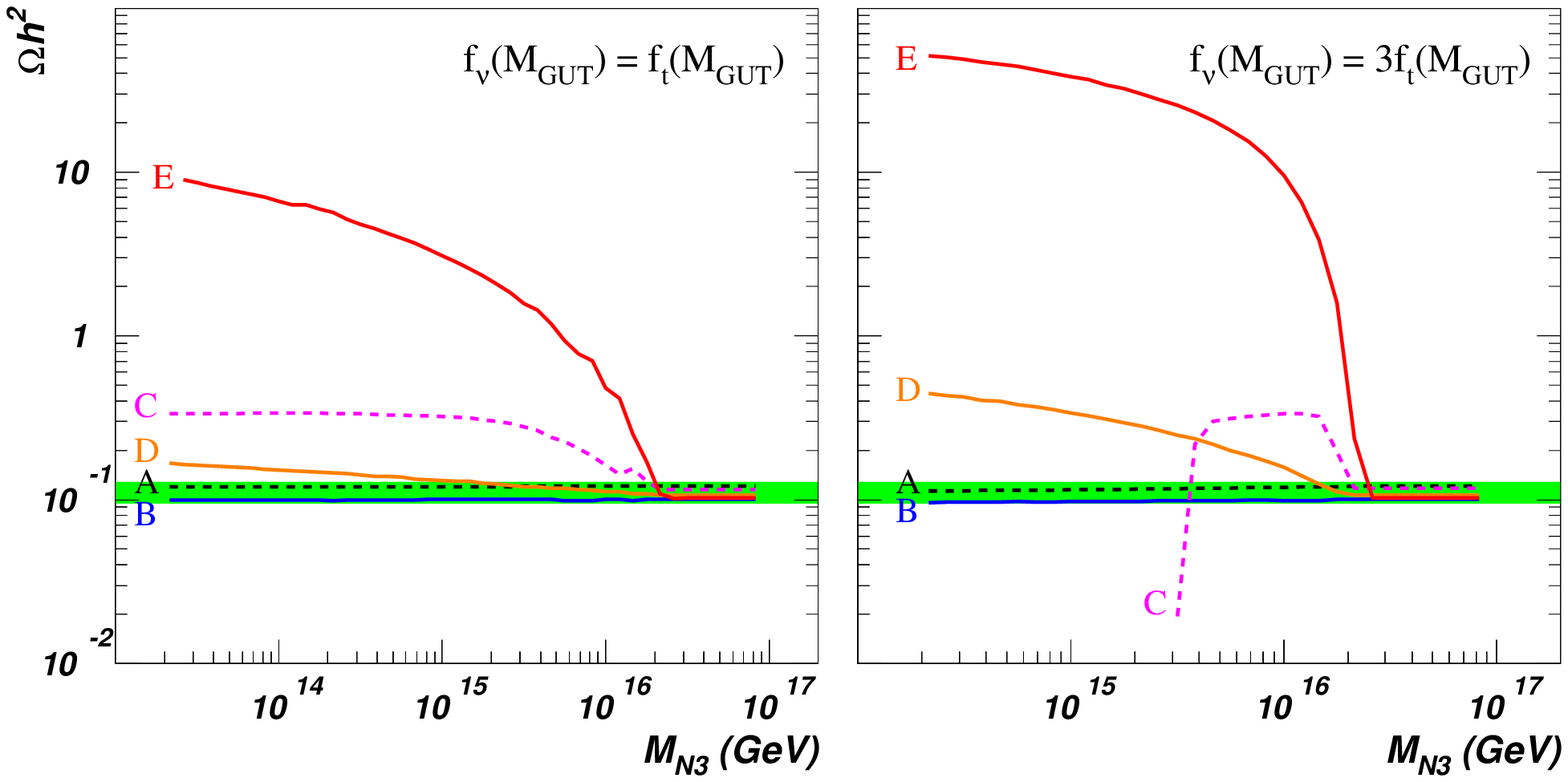,width=14cm,angle=0} 
\caption{\label{fig:rd} 
Neutralino relic density as a function of the GUT-scale value of $M_{N_3}$
for two values of the GUT-scale neutrino Yukawa coupling: $\fnu (M_{GUT})=f_t(M_{GUT})$ and 
$\fnu (M_{GUT})=3f_t(M_{GUT})$.
The labels on the curves refer to the points in Table~\ref{tab:points} and the green bands mark 
the WMAP range of Eq.~(\ref{eq:relic}).
The curves do not extend below the value of $M_{N_3}$ for which the tau-neutrino becomes heavier than 0.3~eV.
}}

To properly account for RHNs, we upgraded ISAJET~\cite{isajet} to full matrix form and added
RGEs for evolution of the RHN Majorana mass matrix $M_N$ above the scale of
decoupling, and a dimension-5 operator $\kappa$ below it~\cite{isajetm}. 
To evaluate the neutralino relic density\footnote{We adopt the conceptually simplest scenario 
in which neutralinos are thermally produced in the standard
$\Lambda$CDM cosmology and make up the DM.}
and direct DM detection
rates we employ the IsaRED~\cite{isared} and IsaRES~\cite{isares} 
subroutines of the IsaTools package. To compute the indirect DM detection rates we use the DarkSUSY~\cite{darksusy} package.
We evaluate the integrated continuum $\gamma$-ray flux above a $E_{\gamma}=1$~GeV threshold.
For positrons and antiprotons, we compute the differential flux at a kinetic energy of 20~GeV, for which
 optimal statistics and signal-to-background ratio are expected at spaceborne antiparticle 
detectors~\cite{antimatter}. For
antideuterons, we compute the average differential flux in the $0.1 < T_{\bar{D}} < 0.25$~GeV range, where 
$T_{\bar{D}}$ is the antideuteron kinetic energy per nucleon~\cite{dbar}.
We performed cross-checks against ISAJET~7.77 with IsaTools as well as
SPheno~2.2.3~\cite{spheno} interfaced with micrOMEGAs~2.0.7~\cite{mo} and found results in good agreement.

The effect of neutrino Yukawa couplings on the neutralino relic density is illustrated in Fig.~\ref{fig:rd}, which shows 
$\Omega_{\tz_1}h^2$ versus the GUT-scale value $M_{N_3}$ for the points in Table~\ref{tab:points}.
The curves do not extend below the value of $M_{N_3}$ for which the tau-neutrino\footnote{Since flavor effects are 
irrelevant for our study, we ignore mixing in both quark and lepton sectors, thus
treating neutrino favor eigenstates as mass eigenstates.}
 becomes heavier than 0.3~eV, in accordance
with the cosmological bound on neutrino masses, \mbox{$\sum m_{\nu} \lesssim 1$~eV}~\cite{wmap}. 
The curves become flat for $M_{N_3}\gtrsim 2\times 10^{16}$~GeV because the right-handed neutrino decouples 
above the GUT scale and has no effect on the spectrum.
We see that, contrary to common belief, neutrino Yukawa couplings {\it significantly impact} the $\tz_1$ relic density. 
In what follows, we provide detailed explanations of this effect and discuss consequences for DM detection rates.

\subsection{Bulk region}
\label{ch:bulk}

\FIGURE{
\epsfig{file=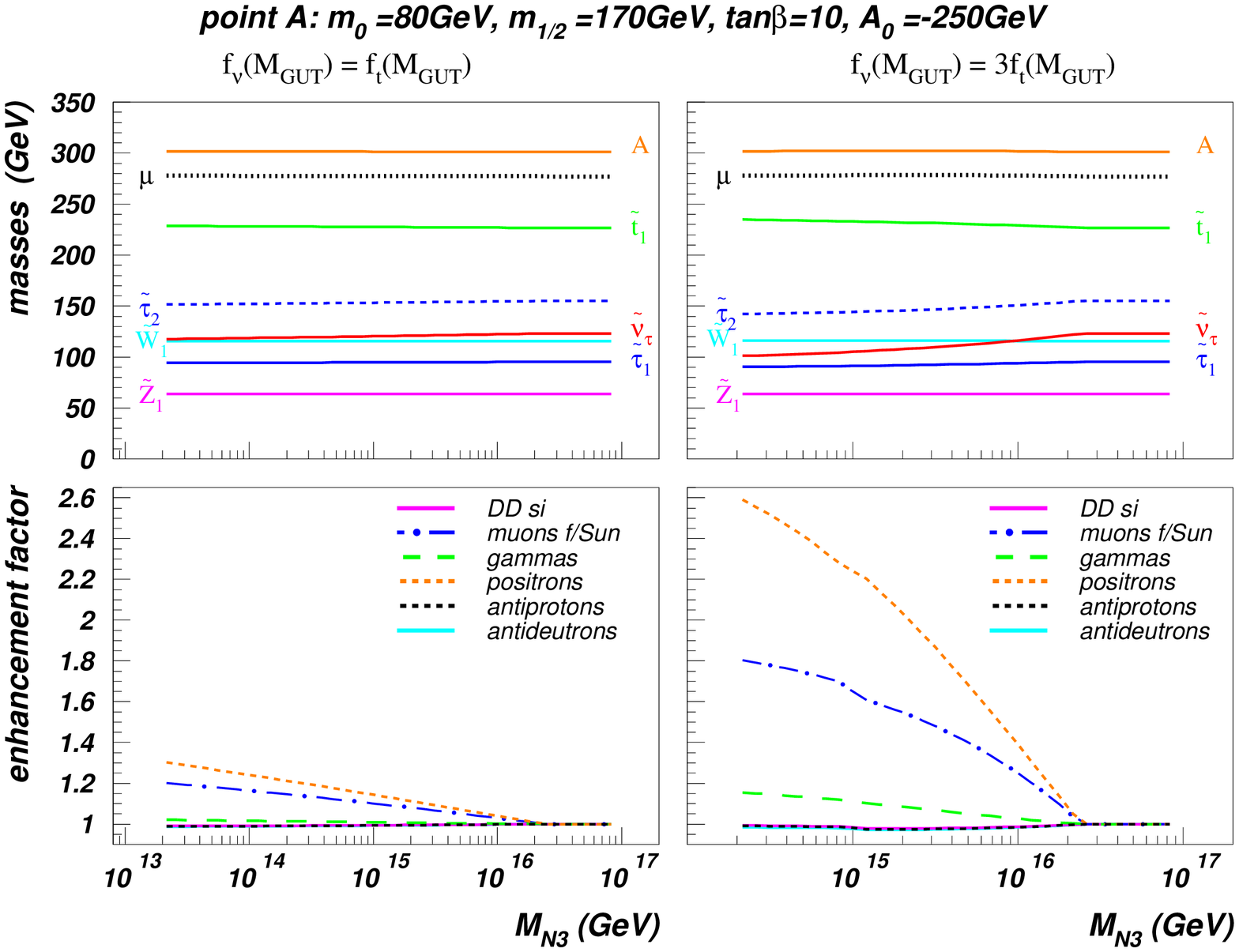,width=15.0cm,angle=0} 
\caption{\label{fig:bulk} 
Some relevant sparticle masses (upper frames) and
enhancement factors, {\it i.e.}, ratio of the rate in mSUGRA+RHN to the corresponding rate in
mSUGRA, for direct and indirect DM detection (lower frames)  
as a function of $M_{N_3}$ at the GUT scale for the benchmark point in the bulk region. The legend is described
in the text. The neutrino Yukawa coupling is $\fnu (M_{GUT})=f_t(M_{GUT})$ in the left frames and 
$\fnu (M_{GUT})=3f_t(M_{GUT})$ in the right frames. 
}}

If the neutralino is bino-like, as in many SUSY models, and sparticles are light ($\sim 100$~GeV) 
then $\tz_1$ can efficiently annihilate into a pair of SM
fermions via a sfermion exchange in the $t-$channel. The dominant process is $\tz_1 \tz_1 \rightarrow l
\bar{l}$, since right-handed sleptons have the largest hypercharge value ($Y(\hat{E^c})=1$). In mSUGRA this
occurs in the bulk region of parameter space which has small values of $m_0,\ m_{1/2},\ A_0$. This region was favored in early
work on mSUGRA, but is largely excluded by the nonobservance of chargino and slepton pair production at
LEP2~\cite{bulk,Baer:1995nc}.

In the upper frames of Fig.~\ref{fig:bulk}, we plot relevant sparticle masses and the $\mu$ parameter versus
the GUT-scale value $M_{N_3}$ for point A of
Table~\ref{tab:points}. We see that the tau-sneutrino gets lighter as $M_{N_3}$ decreases --
a smaller $M_{N_3}$ means a lower seesaw scale and a correspondingly greater downward push from the
 $\fnu^2 X_{\nu}$ term in Eq.~(\ref{eq:ml3}).
 However, due to the smallness of $m_0$ and $A_0$, the overall effect is small --
 up to $\sim 3\%$ for $\fnu (M_{GUT})=f_t(M_{GUT})$ and 
 up to $\sim 15\%$ for $\fnu (M_{GUT})=3f_t(M_{GUT})$.
Because of the universality of GUT-scale SSB boundary conditions, the lightest stau is mostly right-handed and hence
its mass is affected only slightly. On the other hand, $\ttau_2$ is mostly left-handed and experiences changes similarly to $\tnu_{\tau}$.  
The $\fnu^2 \anu$ term in Eq.~(\ref{eq:at}) makes $A_t$ less negative leading to a slightly heavier $\ttop_1$: by
up to $\sim 1\%$ in the left frame and by up to $\sim 3.5\%$ in the right frame.
The rest of the masses are unaffected by the right-handed neutrino. 
As a result the neutralino relic density is barely affected; see curves A in Fig.~\ref{fig:rd}.

The lower frames of Fig.~\ref{fig:bulk} show the effect of changing the right-handed neutrino mass on neutralino DM direct (DD)
and indirect detection (IDD) rates.
It is well-known that the rates are sensitive to the (unknown) DM halo distribution as well as values of hadronic
parameters~\cite{griest,sigmaterm}. 
We define an enhancement factor as the ratio of the rate in mSUGRA+RHN to the corresponding
rate in mSUGRA, so that it is approximately independent of those uncertainties.
We see that as $M_{N_3}$ decreases, the rates for positrons (orange dashed curve), muons from the Sun (blue
dash-dotted curve) and $\gamma$-rays from the Galactic center (green long-dashed curve) increase. 
This is because as $M_{N_3}$ decreases, $\ttau_2$ gets lighter, and neutralino annihilation in $\tau$ lepton pairs is
enhanced with a concomitant enhancement of the above IDD rates. The difference in enhancement factors corresponds to the role
$\tau$ leptons play in these rates: $\tau$'s are the dominant source for positrons and muons (since vector
boson production is kinematically forbidden for the parameter point considered), but contribute little to
gamma-ray production~\cite{griest}. 
For the $\fnu (M_{GUT})=f_t(M_{GUT})$ case, rates for positrons, muons and
$\gamma$-rays increase by up to $30\%,\ 20\%$ and $2\%$ respectively. 
In the case of $\fnu (M_{GUT})=3f_t(M_{GUT})$, the $\ttau_2$ mass is pushed closer to the $\tz_1$ mass
causing larger enhancements of the IDD rates. Rates increase by up to 160\% for positrons, 80\% for muons and
20\% for $\gamma$-rays as shown in the lower right frame of Fig.~\ref{fig:bulk}.
Direct detection rates, that are conventionally represented as the cross-section of spin-independent elastic
neutralino-proton scattering (solid magenta curve), as well as the antiproton (dashed black curve) and
antideuteron (solid turquoise curve) rates 
are unaffected by the $\tz_1\tz_1 \rightarrow \tau \bar{\tau}$ enhancement; their tiny suppression (less than 1\%) is due
to the small change in sbottom mixing caused by radiative corrections.

\subsection{Stau-coannihilation region}
\label{ch:stau}

If the mass of a bino-like neutralino is close to the mass of a more strongly interacting sparticle, 
then $\Omega_{\tz_1}h^2$ is lowered via coannihilation. Usually the lightest stau $\ttau_1$ acts as such a sparticle and
rapid reactions $\tz_1\ttau_1 \rightarrow X$ and $\ttau_1 \ttau_1 \rightarrow X$ 
(where $X$ denotes any allowed final state of SM or Higgs particles),
in the early universe make the $\tz_1$ relic density consistent with WMAP data. 
In mSUGRA this region appears at small $m_0$ and small to medium $m_{1/2}$ values~\cite{dm:stau,isared}.

The situation is similar to that in the bulk region -- $\tnu_{\tau}$ and $\ttau_2$ get
lighter as $M_{N_3}$ decreases, while the rest of the sparticles are unaffected. However, the diminution is
smaller ($\lesssim 1\%$) because $A_0=0$ and $m_{1/2} > m_0$ reduces the relative role of the $\fnu$ terms in
the RGE evolution. Again, due to the right-handedness of $\ttau_1$, its mass and concomitantly $\Omega_{\tz_1}h^2$ remain
unchanged -- see curves B in Fig.~\ref{fig:rd}.


The smaller change in the $\ttau_2$ mass leads to a smaller increase in the $\tz_1\tz_1 \rightarrow \tau\bar{\tau}$ rate, which in
turn yields a smaller enhancement for muon and positron IDD rates 
(\mbox{$\lesssim 2\%$} for $\fnu=f_t$ at the GUT scale and $\lesssim 12\%$ for $\fnu=3f_t$)
as compared to the bulk region. 
Also, in this region, neutralino pair annihilation into vector bosons is allowed since
\mbox{$m_{\tz_1}\simeq 0.5 m_{1/2} > M_Z$}. This 
process dominates over $\tz_1$ pair annihilation into $\tau$ leptons as the source of muons and positrons and is unaffected by the 
RHN mass; this further reduces the sensitivity of these IDD rates to changes in the $\ttau_2$ mass. 
The effect of varying $M_{N_3}$ on the DD and the other IDD rates is the same as in 
the bulk region and is of the same order.

\subsection{Stop-coannihilation region}
\label{ch:stop}

Another possibility for coannihilation is with the scalar top of mass $m_{\ttop_1} \simeq
m_{\tz_1}$. In mSUGRA this region is located at small $m_0$ and $m_{1/2}$ and is
characterized by a large value of $A_0$~\cite{dm:stop}.

For $|A_0|\gg m_0,\ m_{1/2}$, the A-terms dominate the RGE evolution. The large negative
$\fnu^2 \anu$ term pushes $|A_t|$ and $|\atau|$ to smaller values at the weak scale, which reduces the L-R mixing
in stops and staus. The smaller $|A_t|$ contribution to $X_t$ also reduces the downward push by the $f_t^2 X_t$ term in $m^2_{\ttop}$ 
evolution. A combination of these effects increases $m_{\ttop_1}$ away from the neutralino mass,
thus shutting off the coannihilation mechanism.
Also, the large $X_{\nu}$ term in Eq.~(\ref{eq:ml3}) drives the masses of the
left-handed sleptons $\tnu_{\tau}$ and $\ttau_2$ to smaller values; see Fig.~\ref{fig:stop}.
For $\fnu (M_{GUT})=f_t(M_{GUT})$, the growing stop mass
causes the neutralino relic density to rapidly increase above the WMAP-preferred range to $\Omega_{\tz_1}h^2 \simeq 0.3$ for $M_{N_3}
\lesssim 2\times 10^{15}$~GeV; see curve C in the left frame of Fig.~\ref{fig:rd}.

The dependence on $M_{N_3}$ is more significant for the $\fnu (M_{GUT})=3f_t(M_{GUT})$ case, shown in the upper right frame of
Fig.~\ref{fig:stop}. Such a large neutrino Yukawa coupling increases the importance of the $\fnu$ terms in the RGEs. The stop
mass is pushed to even higher values, while $m^2_{L_3}$ falls faster and causes stau masses to decrease by $\sim
23\%$; staus also undergo an identity flip at $M_{N_3}\simeq 10^{16}$~GeV -- $\ttau_1$ changes from a
right-handed to dominantly a left-handed state. But most importantly, the tau-sneutrino gets lighter
(with mass $\simeq m_{L_3}$) and becomes the NLSP with almost the same mass as the neutralino for
$M_{N_3}\lesssim 4.5\times 10^{16}$~GeV. This opens up the $\tz_1 - \tnu_{\tau}$ coannihilation channels that rapidly decrease
$\Omega_{\tz_1}h^2$ down to and then below the WMAP range, as shown by curve C in the left frame of
Fig.~\ref{fig:rd}. Below $M_{N_3}\simeq 3\times 10^{16}$~GeV, the $\tnu_{\tau}$ gets lighter than the
$\tz_1$ and the parameter space closes.

\FIGURE{
\epsfig{file=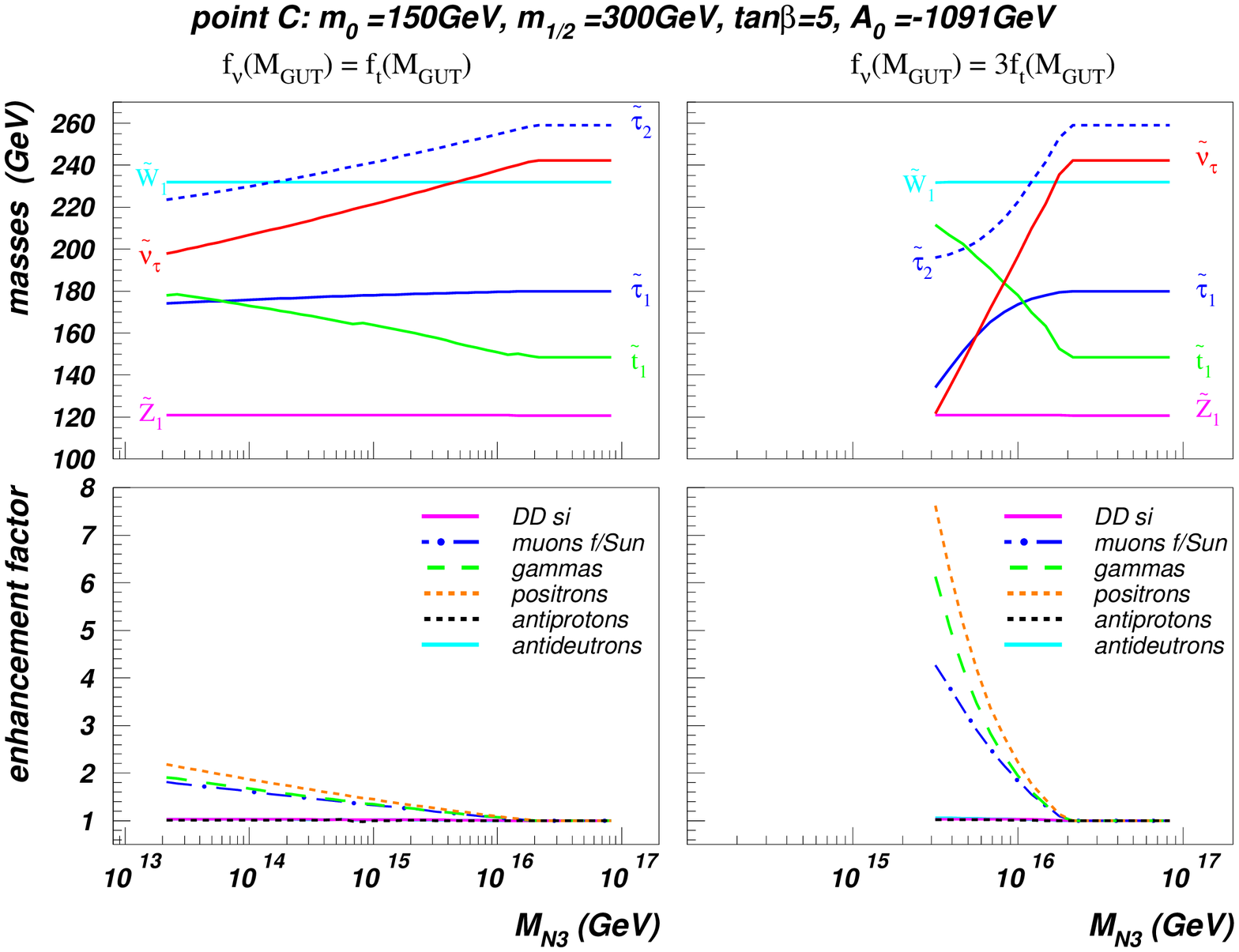,width=15.0cm,angle=0} 
\caption{\label{fig:stop} 
Similar to Fig.~\ref{fig:bulk}, but for the benchmark point in the stop-coannihilation region.
}}

As $M_{N_3}$ is dialed to smaller values, tau-slepton masses get smaller, leading to enhanced $\tz_1$ pair annihilation into $\tau$-leptons in the $t-$channel. 
Since $\tau$'s are one of the primary sources for muons and positrons, the corresponding IDD rates increase
by up to 100\% and 150\% respectively in the $\fnu (M_{GUT})=f_t(M_{GUT})$ case. Similarly, $\gamma$-rays originate from $\tau$'s and therefore
their IDD rate changes by up to 130\%.
In the $\fnu (M_{GUT})=3f_t(M_{GUT})$ case,  significantly lighter staus cause the IDD rates to increase by up to $300\% - 700\%$.

Antiprotons and antideuterons are mainly produced in the hadronization of heavy quark flavors (bottom and top)
present in the products of $\tz_1$ pair annihilation. In general, increasing the $\ttop_1$ mass suppresses annihilation rates
into $t \bar{t}$. However, since neutralinos in the halo are essentially at rest, the process $\tz_1 \tz_1 \rightarrow t
\bar{t}$ is kinematically forbidden for the point considered. Sbottoms, on the other hand, are very heavy and
do not significantly contribute to $\tz_1 \tz_1 \rightarrow b\bar{b}$. As a result, IDD rates for antiprotons
and antideuterons change by less than 1\% as $M_{N_3}$ is varied.

\subsection{Higgs funnel}

\FIGURE{
\epsfig{file=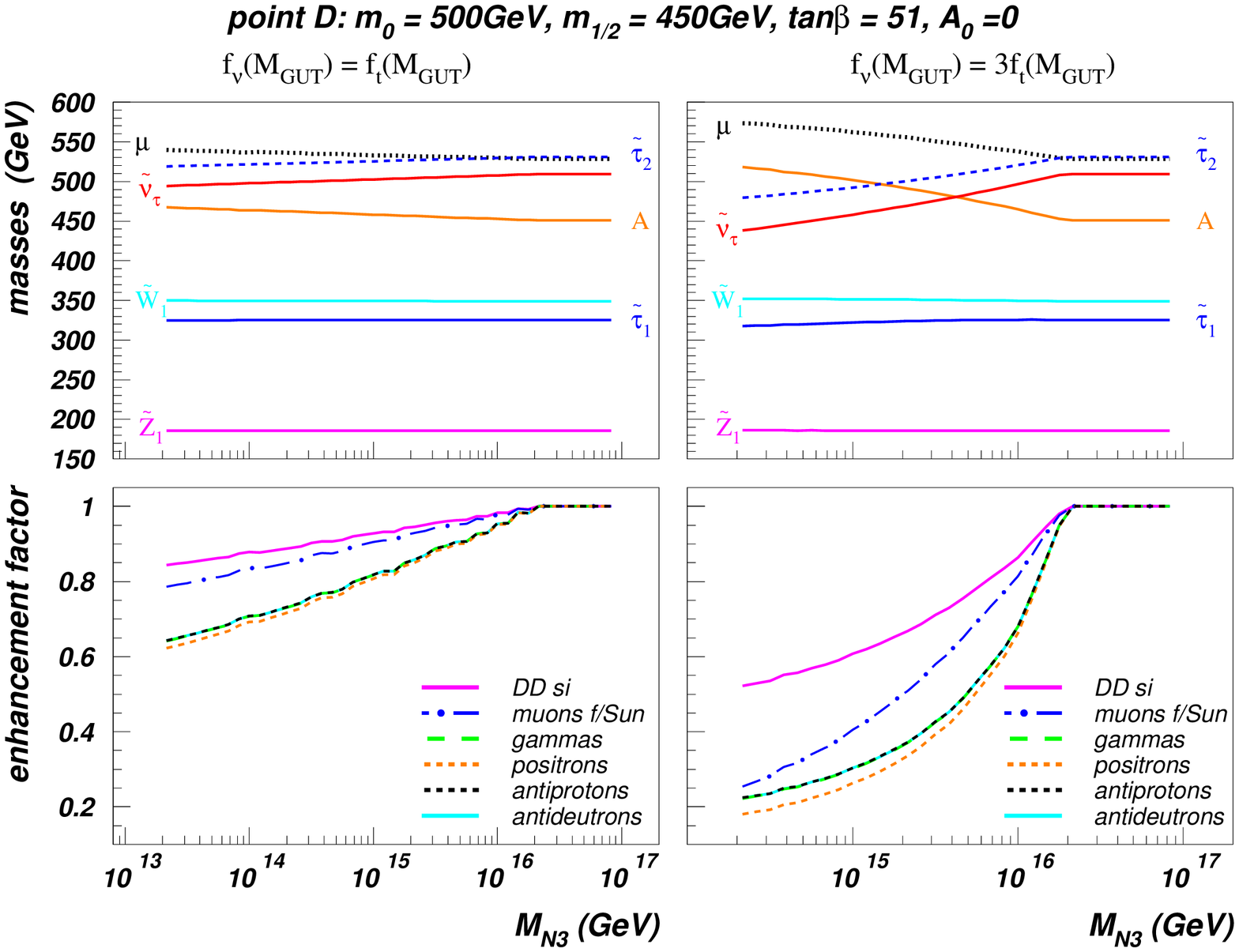,width=15.0cm,angle=0}
\caption{\label{fig:afunnel}
Similar to Fig.~\ref{fig:bulk}, but for the benchmark point in the $A$-funnel.
}}

If $2m_{\tz_1}\simeq m_A$, neutralino annihilation through the $A$ (and $H$) Higgs boson in the
$s-$channel are resonantly enhanced. Since the $A$-width can be very large ($\Gamma_A \sim 10-50$~GeV), an exact
equality in the mass relation is not necessary to achieve the near-resonance enhancement.
This WMAP-compatible region is known as the $A$-funnel~\cite{Afunnel} and occurs in mSUGRA at medium $m_0$ and large
$\tan\beta$.

Increasing $m_0$ increases the downward push by the $\fnu^2 X_{\nu}$ term in Eq.~(\ref{eq:ml3}),
leading to a lighter $\tnu_{\tau}$ and $\ttau_2$. Since $A_0=0$, the L-R mixing is small, $\ttau_1$ is
right-handed and therefore almost unaffected. Similarly, the $\fnu^2 X_{\nu}$ term in Eq.~(\ref{eq:mhu}) 
pushes $m^2_{H_u}$ to more negative values resulting in larger $|\mu|$ and $m_A$ values. 
This is borne out in the upper frames of Fig.~\ref{fig:afunnel}.
Larger $|\mu|$ translates into heavier ($\lesssim 7\%$) higgsino-like $\tz_3,\ \tz_4,\ \tw_2$ states.
Increasing the mass of the CP-odd Higgs boson $A$ pushes it away from the resonance resulting in
a reduction of the $\tz_1$ annihilation rate and thus larger relic density. From curves D in
Fig.~\ref{fig:rd}, we see that $\Omega_{\tz_1}h^2$ is above the WMAP range for $M_{N_3}\lesssim 10^{15}$~GeV (left
frame); a larger neutrino Yukawa coupling causes $m_A$ to grow faster and $\Omega_{\tz_1}h^2$ exceeds the WMAP
range for $M_{N_3}\lesssim 10^{16}$~GeV (right frame).

The same neutralino annihilation mechanism $\tz_1\tz_1 \rightarrow A \rightarrow b\bar{b} / \tau \bar{\tau}$ that reduces
the relic density plays a primary role in DM halo annihilation. Consequently, its reduction has a significant
effect on IDD rates. Increasing the mass of the $A$ decreases the production of $b$-quarks in the $\tz_1$ halo, thus
reducing antiproton and antideuteron fluxes; the flux of $\gamma$-rays produced via the $b \rightarrow
\pi^0 \rightarrow \gamma$ chain also experiences similar suppression. 
From the lower frames of Fig.~\ref{fig:afunnel}, we see that the effect can reach $\sim 40\%$ if 
$\fnu(M_{GUT})=f_t(M_{GUT})$ and $\sim 75\%$ if $\fnu(M_{GUT})=3f_t(M_{GUT})$.

The exchange of $h$ and $H$ bosons are important for neutralino-proton elastic scattering. 
Therefore, increasing $m_H$ leads to a suppression of DD rates by up to about 15\% in the left frame and up to
$\sim 50\%$ in the right frame.
The spin-dependent $\tz_1 - p$ elastic cross section suffers a similar suppression thus lowering the neutralino capture rate
in the Sun. This effect is coupled with a reduction of the neutralino annihilation cross section. Together, the solar muon flux
is reduced by $\sim 20\%$ for $\fnu(M_{GUT})=f_t(M_{GUT})$ and by $\sim 75\%$ for 
$\fnu(M_{GUT})=3f_t(M_{GUT})$.

\subsection{Hyperbolic branch/Focus point region}

\FIGURE{
\epsfig{file=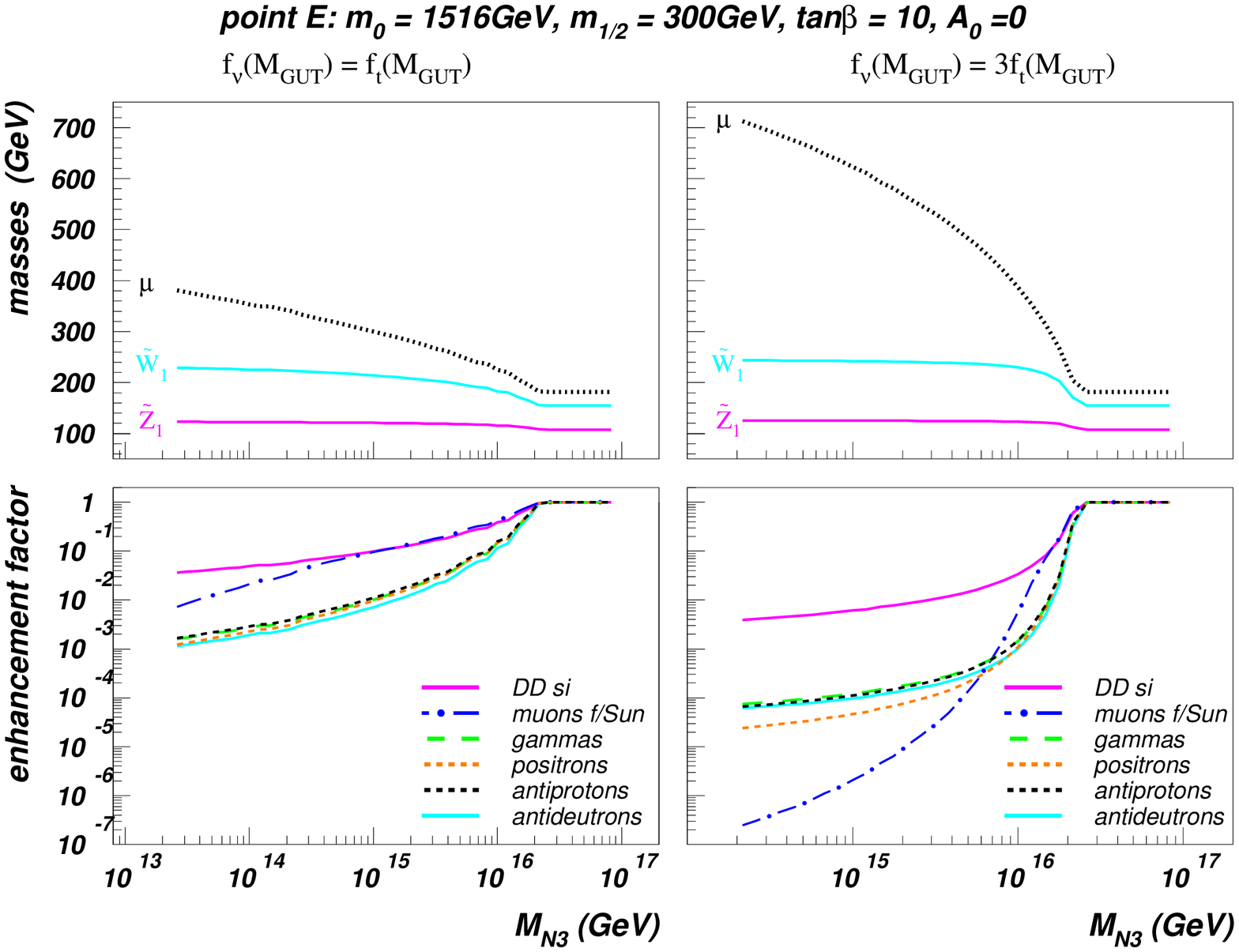,width=15.0cm,angle=0} 
\caption{\label{fig:fp} 
Similar to Fig.~\ref{fig:bulk}, but for the benchmark point in the HB/FP region.
}}

If $\mu$ is sufficiently small 
the lightest neutralino develops a sizable higgsino
component, which enhances $\tz_1$ annihilation to $WW,\ ZZ$ and $Zh$ in the early universe and brings
$\Omega_{\tz_1}h^2$ in accord with Eq.~(\ref{eq:relic}). 
In mSUGRA this is realized in the hyperbolic branch or focus point region located at very large $m_0$ along the edge
of the no-REWSB region where $\mu^2$ becomes negative~\cite{dm:fp}.

The presence of a large neutrino Yukawa coupling in Eq.~(\ref{eq:mhu}) results in a
larger absolute value for $m^2_{H_u}$ at the weak scale and a corresponding increase in $\mu$. 
This means that the neutralino becomes increasingly bino-like and its relic density rapidly grows
beyond the WMAP range, as shown by curves E in Fig.~\ref{fig:rd}.
The lightest neutralino mass increases by about 15\%, while the lightest chargino mass grows by 50\% and changes composition
from higgsino to wino; see the upper frames of Fig.~\ref{fig:fp}. 
The masses of heavier charginos and neutralinos also increase by 50\% to 100\%, as $|\mu|$ increases. 
As $|m^2_{H_u}|$ increases, the $A$-mass also increases by up to $2.5\%$ (up to $10\%$ for the larger $\fnu$ case).
As in Section~\ref{ch:stop}, decreasing $m^2_{L_3}$ causes the mass of the $\tnu_{\tau}$ to drop by
$\sim 3\%$ (up to $\sim 15\%$ for the larger $\fnu$ case) and the $\ttau_1$ to change its composition to left-handed state at $M_{N_3}\lesssim 2\times
10^{15}$~GeV ($M_{N_3}\lesssim 10^{15}$~GeV for the larger $\fnu$ case) with an accompanying reduction in mass.
However, due to the large $m_0$ value, all sfermions are very heavy and are not relevant for DM detection.

The decrease of the higgsino composition of $\tz_1$ has a significant effect on DM detection -- as $M_{N_3}$ decreases,
all the rates drop by {\it several orders of magnitude}. 
The lower higgsino content reduces
annihilation into vector bosons and suppresses the overall $\tz_1$ annihilation rate. As a result $\gamma$-ray and
antimatter fluxes from neutralino annihilations in the galactic halo fall by a factor of $10^3$ ($\sim 10^4$ in the right frame). 
DD rates decrease by up to about two (three) orders of magnitude because the coupling of $\tz_1$ to higgs bosons is diminished.
The muon flux is also reduced due to a decreasing neutralino capture rate and weakening $\tz_1$ pair
annihilation: the overall reduction reaches two orders of magnitude for $\fnu (M_{GUT})=f_t(M_{GUT})$ and more than
six orders of magnitude for $\fnu (M_{GUT})=3f_t(M_{GUT})$.

\section{Discussion and Conclusions}
\label{disc}



So far our discussion was confined to particular values of $m_0,\ m_{1/2},\ A_0$, but neutrino Yukawa
couplings generally affect the WMAP-allowed regions in mSUGRA parameter space.
To illustrate this we performed a random scan in $(m_0,\ m_{1/2},\ M_{N_3})$ and plotted results in the
conventional $(m_0,\ m_{1/2})$ plane after marginalizing over $M_{N_3}$. 
In the green regions of Fig.~\ref{fig:rscan}, the neutralino relic density lies within the WMAP
$2\sigma$-range given by Eq.~(\ref{eq:relic}). We do not show points that are excluded by the LEP2
chargino bound $m_{\tw_1} > 103.5$~GeV~\cite{lep2}.
For comparison, we superimposed the WMAP-allowed regions in mSUGRA as blue crosses. 
The stau-coannihilation region is virtually unaffected by the neutrino Yukawa coupling -- as explained in
Section~\ref{ch:stau}.
On the other hand, the $\mu$ parameter in the HB/FP region gets larger for smaller $M_{N_3}$, thus postponing the
breakdown of the REWSB mechanism to larger $m_0$ values.
This makes the HB/FP region very sensitive to the neutrino Yukawa coupling -- 
depending on the value of $M_{N_3}$ it can move to the right by up to $\sim 500$~GeV, and even more
if $\fnu=3f_t$ at the GUT scale.
This shift of the HB/FP region to larger $m_0$ values makes sfermions heavier and can have a sizable effect on 
collider signatures.

\FIGURE{
\epsfig{file=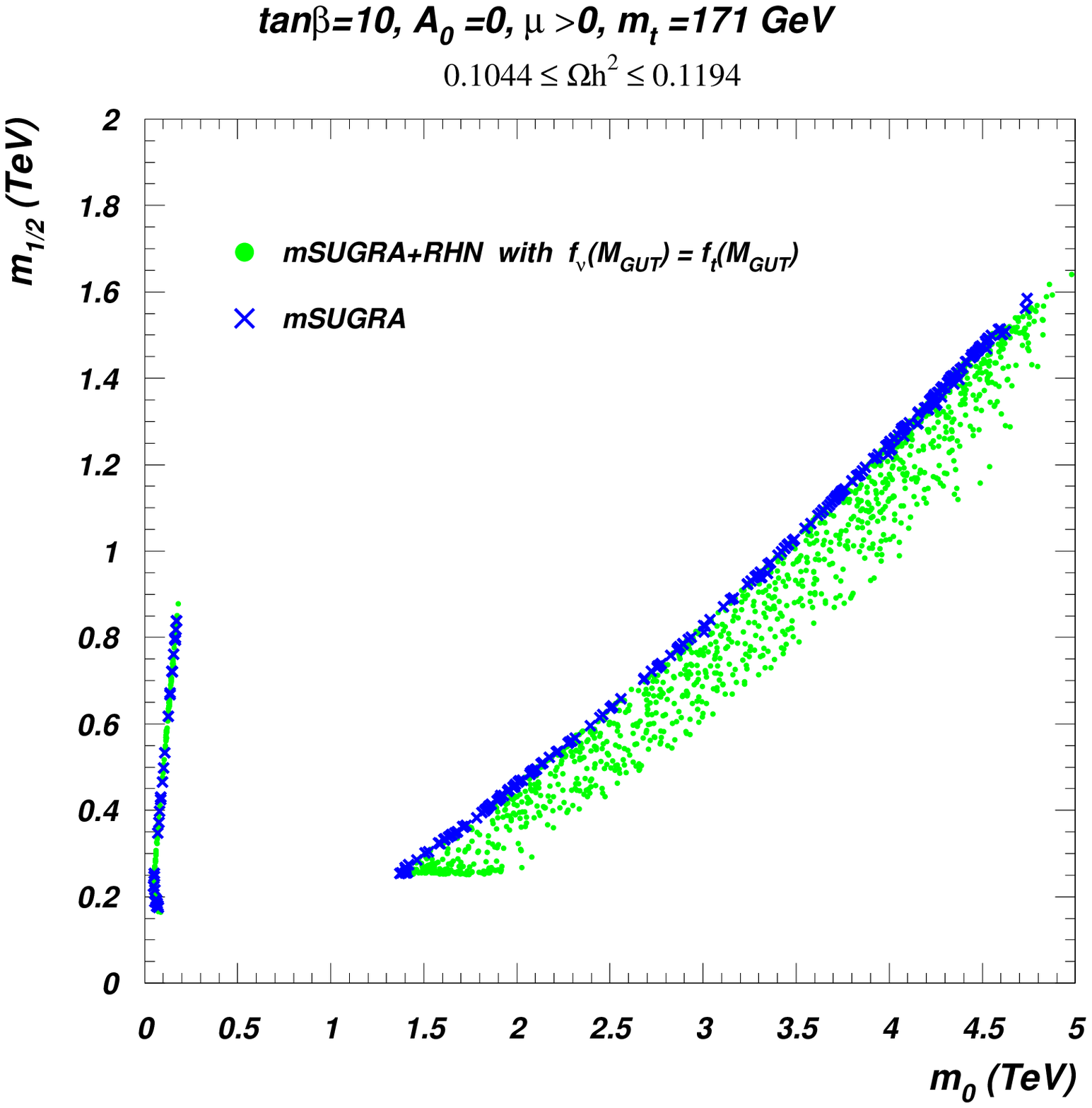,width=10.0cm,angle=0} 
\caption{\label{fig:rscan} 
Regions compatible with the dark matter relic density in the $(m_0, \ m_{1/2})$ plane 
for the mSUGRA+RHN model with $f_{\nu}(M_{GUT})=f_t(M_{GUT})$, $A_0=0,\ \mu >0$ and $m_t=171$~GeV. 
The green points have a neutralino relic density within the WMAP
$2\sigma$-range (Eq.~\ref{eq:relic}).
The right-handed neutrino mass $M_{N_3}$ is randomly varied between $10^{13}$~GeV and $M_{GUT}\simeq 2\times10^{16}$~GeV,
with the requirement that tau-neutrino be lighter than 0.3~eV.
The blue crosses mark the WMAP-allowed regions in the mSUGRA model. Note the significant expansion of the parameter
space when neutrino masses are accounted for.
}}

If the mechanism that lowers the relic density also affects DM detection rates, then rates 
in the WMAP allowed regions may be qualitatively unaltered from mSUGRA expectations, as is the case
in the HB/FP region where both the relic density and detection rates are tied to the higgsino component of $\tz_1$.
This is because neutrino Yukawa coupling effects can be compensated by moderate shifts in the space of fundamental parameters. 
However, when the mechanisms dictating the relic density and detection rates are different, and 
such a compensation is not possible, one can expect sizeble variations in DM detection rates. 
In the bulk and stau-coannihilation regions, the relic density remains unaffected by dialing $M_{N_3}$, but the rates can 
change appreciably -- see Sections~\ref{ch:bulk} and~\ref{ch:stau}. Since the specific realization
of the seesaw mechanism is unknown, it is an additional uncertainty in the DM detection rates.

Although we presented our analysis of the effects of neutrino Yukawa couplings in a mSUGRA+RHN framework, 
the effects appear in any SUSY-seesaw model in which RHNs with large Yukawa couplings decouple below the GUT scale.
Moreover, the effect can be even larger in a non-mSUGRA framework due to different sparticle properties~\cite{wtn}.
For instance, the right-handedness of the $\ttau_1$ that limited the effect of neutrinos in mSUGRA (especially in the 
stau-coannihilation region) does not hold in scenarios in which one or both
higgs mass parameters are non-universal at the GUT scale, as in the models of Ref.~\cite{nuhm}.
Here, $\ttau_1$ is dominantly $\ttau_L$ and thus more susceptible to neutrino Yukawa couplings.
Also, the $A$-funnel appears in mSUGRA only at $\tan\beta$ where bottom and tau Yukawa
couplings are large, thus dampening the effect of $\fnu$ on RGE evolution.  
In the models of Ref.~\cite{nuhm}, the $A$-funnel can also occur
at small and medium  $\tan\beta$ values and we expect even larger changes of DM detection rates with variations of
the RHN mass.

We emphasize that although we used $SO(10)$ models to estimate the size of the neutrino Yukawa coupling at the GUT scale, 
our results are not limited to models that employ this gauge group. 
In fact, in $SU(5)$-based SUSY GUTs, where the right-handed neutrinos are
singlets, $\bfn$ is not correlated with the other Yukawas and can be even larger than we considered. Requiring
perturbativity, neutrino Yukawa couplings are limited to \mbox{$\bfn \lesssim 15 \bf{f}_u$~\cite{gunif}}.

It is also worth keeping in mind that many mechanisms other than the type~I seesaw have been suggested for the generation
of neutrino masses. For example, in a
double seesaw~\cite{seesaw1}, one postulates an $SO(10)$ singlet
with mass $M_S$ in addition to a right-handed neutrino $N$, and finds that the light neutrino mass takes the form 
$m_{\nu} \sim M_S v^2_u \fnu^2 / M_{N_3}^2$. Since $M_S \ll M_{N_3}$,
 $M_{N_3}$ would be considerably smaller than in a type~I seesaw: if
$M_S \sim 1$~TeV, the RHN could be as light as $10^8$~GeV. In this case, changes in DM observables could be
significantly larger.

In summary, motivated by the concrete evidence that neutrinos are massive, 
we examined the effect of neutrino Yukawa couplings on the neutralino relic density and dark matter detection
rates in SUSY seesaw models.
We found that, contrary to common belief, neutrino Yukawa couplings
can significantly affect the relic density. Effects are most prominent in regions
of parameter space where SUSY-breaking slepton masses and/or trilinear couplings are large. 
Such conditions are satisfied, for example, in the $A$-funnel, focus point and stop-coannihilation regions
of mSUGRA. Here the neutralino relic density can change by over an order of magnitude due to the effect of neutrino
Yukawa couplings; see Fig.~\ref{fig:rd}. We also showed that DM detection rates can be changed by up to several orders
of magnitude in the focus point region and by factors of a few to ten in the other regions; see Figs.~\ref{fig:bulk}--\ref{fig:fp}.

\section*{Acknowledgments}
We thank H.~Baer, A.~Belyaev and A.~Soleimani for useful inputs,
and the Kavli Institute for Theoretical Physics at the University of California,
Santa Barbara for its support and hospitality.
This work was supported by the DoE under Grant Nos. DE-FG02-95ER40896 and DE-FG02-04ER41308,
by the NSF under Grant Nos. PHY-0544278 and PHY05-51164,
 and by the Wisconsin Alumni Research Foundation.


%
\end{document}